\documentclass[useAMS,usenatbib]{mn2e}
\usepackage{graphicx}

\begin{document}

\title[Confronting Cosmology and New Physics with Fundamental Constants]{Confronting Cosmology and New Physics with Fundamental Constants}
\author[Rodger I. Thompson]{Rodger I. Thompson$^{1}$\thanks{E-mail:
rit@email.arizona.edu (RIT)}\\
$^{1}$Steward Observatory, University of Arizona, Tucson, AZ 85721, USA\\}

\pagerange{\pageref{firstpage}--\pageref{lastpage}} \pubyear{2013}

\maketitle

\label{firstpage}

\begin{abstract}
The values of the fundamental constants such as $\mu = m_P/m_e$, the proton to electron mass 
ratio and $\alpha$, the fine structure constant, are sensitive to the product $\sqrt{\zeta_x^2(w+1)}$
where $\zeta_x$ is a coupling constant between a rolling scalar field responsible for the 
acceleration of the expansion of the universe and the electromagnetic field with x standing
for either $\mu$ or $\alpha$. The dark energy equation of state $w$ can assume values different 
than $-1$ in cosmologies where the acceleration of the expansion is due to a scalar field. 
In this case the value of both $\mu$ and $\alpha$ changes with time. 
The values of the fundamental constants, therefore, monitor the equation of state
and are a valuable tool for determining $w$ as a function of redshift.
In fact the rolling of the fundamental constants is one of the few definitive discriminators
between acceleration due to a cosmological constant and acceleration due to a
quintessence rolling scalar field. $w$ is often given in parameterized form for comparison with 
observations.  In this manuscript the predicted evolution of $\mu$, is calculated for a 
range of parameterized equation of state models and compared to the observational constraints
on $\Delta \mu / \mu$. We find that the current limits on $\Delta \mu / \mu$ place 
significant constraints on linear equation of state models and on thawing models where
$w$ deviates from $-1$ at late times.  They also constrain non-dynamical models that
have a constant $w$ not equal to $-1$. These constraints are an important compliment to
geometric tests of $w$ in that geometric tests are sensitive to the evolution of the 
universe before the  epoch of observation while fundamental constants are sensitive to the evolution
of the universe after the observational epoch. Recent low redshift radio limits on $\Delta \mu / \mu$
provide the most significant constraints on the late time evolution of $w$. 
\end{abstract}

\begin{keywords}
(cosmology:) cosmological parameters -- dark energy -- theory -- early universe 
\end{keywords}

\section{Introduction} \label{s-intro} 

The apparent acceleration of the expansion of the universe (\citet{rie98} and \citet{per99})
has spurred considerations of new physics and cosmologies beyond the standard model of
physics and the $\Lambda$CDM cosmology. Since the fundamental constants such as the 
electromagnetic fine structure constant $\alpha = \frac{e^2}{\hbar c}$, the proton to
electron mass ratio $\mu = \frac{m_p}{m_e}$ and the gravitational fine structure constant
$\frac{G m_p m_e}{\hbar c}$ determine the quantitative nature of physics it is natural
to check for time variations of these constants as an indicator of physics and cosmologies 
beyond the standard models. Rolling constants are one of the few definitive predictions
of quintessence cosmologies that separate them from standard general relativity cosmological
constant cosmologies.
 
\section{Observational Constraints} \label{s-obcon}

\begin{figure}
  \vspace{70pt}
\resizebox{\textwidth}{!}{\includegraphics[0in,0in][14.in,3.in]{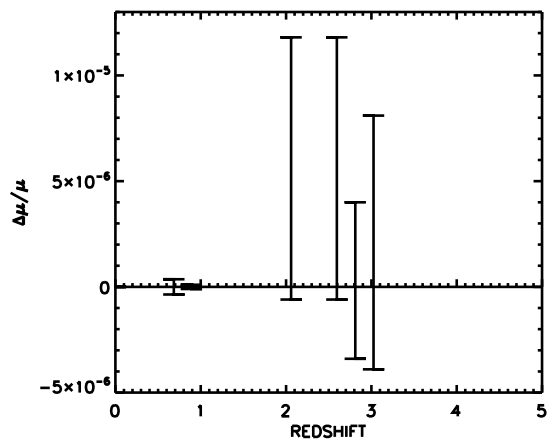}}
  \caption{Observational constraints on $\Delta \mu / \mu$ from radio ($z < 1$)
and optical ($z > 1$) observations.  All constraints are at the $1 \sigma$ 
level except the radio constraint at $z=0.6847$ which is $3 \sigma$.. The radio constraint at 
$z=0.8858$ is difficult to see at this scale but its value is $\pm 1 \times 10^{-7}$.} \label{fig-err}
\end{figure}

The observational constraints on $\Delta \mu / \mu$ are shown in Figure ~\ref{fig-err} at 
the observed redshifts. Table 1 gives the redshifts, constraints, accuracy
and references for the observations.
\begin{table}\label{tab-con}
 \begin{minipage}{120mm}
\begin{tabular}{llllll}
\hline
Object & Redshift  & $\Delta \mu / \mu$ & error  & Acc. & Ref.\\
\hline
Q0347-383 & $3.0249$ & $2.1 \times 10^{-6}$ & $\pm 6. \times 10^{-6}$ & $1\sigma$ & (a)\\
Q0405-443 & $2.5974$ & $10.1 \times 10^{-6}$ & $\pm 6.2 \times 10^{-6}$ & $1\sigma$ & (b)\\
Q0528-250 & $2.811$ & $3.0 \times 10^{-7}$ & $\pm 3.7 \times 10^{-6}$ & $1\sigma$ & (c)\\ 
J2123-005 & $2.059$ & $5.6 \times 10^{-6}$ & $\pm 6.2 \times 10^{-6}$ & $1\sigma$ & (d)\\
PKS1830-211 & $0.88582$ & $0.0$ & $\pm 1.0 \times 10^{-7}$ & $1\sigma$ & (e)\\
B0218+357 & $0.6847$ & $0.0$ & $\pm 3.6 \times 10^{-7}$ & $3\sigma$ & (f)\\
\hline
\end{tabular}
\begin{verse}
Table 1. Observational constraints used in this analysis. \\
 References: (a) \citet{wen08}, (b) \citet{kin09}, \\ 
(c) \citet{kin11}, (d) \citet{mal10}, (e) \citet{bag13}, \\
(f) \citet{kan11}
\end{verse} 
\end{minipage}
\end{table}
At the scale of the figure the constraint at 
$z = 0.88582$ is difficult to see but its value is $\pm 1 \times 10^{-7}$. These constraints 
are the tightest published bounds for an object which may have several
published constraints.  The constraints
at redshifts less than 1 are both radio observations while the remainder are from
optical observations.

\section{Relation between $w$ and $\mu$} \label{s-wmu}

The standard cosmology with a cosmological constant $\Lambda$ predicts a very simple form
of the dark energy equation of state, $w = -1$. Long standing difficulties in 
equating $\Lambda$ with the particle physics vacuum energy \citep{wei89} has led to the 
consideration of other cosmologies with different values of $w$ that can also evolve with
time. Many of these cosmologies invoke a rolling scalar field $\phi$ that also couples with
the electromagnetic field. For cosmologies where a single rolling scalar field couples with
both the gravitational and electromagnetic fields \citet{thm12}, utilizing the work of 
\citet{nun04}, \citet{ave06} and \citet{dut11}, showed that the change in $\mu$ is related
to the EoS $w$ and the coupling $\zeta_{\mu}$ by 
\begin{equation} \label{eq-con}
(w+1)\zeta_{\mu}^2 = \frac{(\mu'/\mu)^2}{3 \Omega_{\phi}}
\end{equation}
where ' indicates the derivative with respect to the log of the scale factor $a$, $ln(a)$.
The evolution of $\mu$ is given by the integral
\begin{equation} \label{eq-muint}
\frac{\Delta\mu}{\mu} =\zeta_{\mu}\int^{a}_{1}\sqrt{3\Omega_{\phi}(x)(w(x)+1)}x^{-1}dx
\end{equation}
which can be numerically integrated \citep{thm12}.  
When $w$ is very close to $-1$ the ratio of the dark energy density to the critical density
is well approximated by
\begin{equation} \label{eq-omg}
\Omega_{\phi} = [1+(\Omega_{\phi 0}^{-1} - 1)a^{-3}]^{-1}
\end{equation}
When the value of $w$ is significantly different than $-1$, however, we use the
more accurate form
\begin{equation}\label{eq-omgc}
\Omega_{\phi}(a) = [1+(\Omega_{\phi 0}^{-1} - 1)a^{-3} exp(3\int^{a}_{1}
\frac{(1+w(x))}{3}dx)]^{-1}
\end{equation}

For any given equation of state $w$
we can then predict the change in $\mu$ as a function of the scale factor $a$ under the
assumption that \emph{$w$ evolves due to a rolling scalar field that is coupled to $\mu$
with a coupling constant given by $\zeta_{\mu}$}. Also note that both equation~\ref{eq-con}
and equation~\ref{eq-muint} indicate that \emph{even a stationary value of $w$ can cause 
$\mu$ to vary if the value is different from $-1$}.  The same statement is true for $\alpha$.

In this work we examine the evolution of $\mu$ for some common paramaterizations of the
EOS.  As a benchmark we utilize a minimum expected value of $\zeta_{\mu}$ of $4 \times 10^{-6}$.
This value accepts the minimum expected value for $\zeta_{\alpha}$ of $10^{-7}$ from 
\citet{nun04} multiplied by the expected ratio of $\Delta \mu /\mu$ to $\Delta \alpha/
\alpha$ of 40-50 \citep{ave06}. Note that the minimum $\zeta_{\alpha}$ is calculated
by assuming that the reported  $\Delta \alpha/ \alpha \approx 10^{-5}$ \cite{web11} is real.
If it is not then $\zeta_{\mu}$ has no definite lower bound.

\section{Equations of State} \label{s-eos}

Several forms of the equation of state $w$ are listed in \citet{def12}, herinafter DNT, 
which include the
CPL or Chevallier-Polarski-Linder parameterization (\citet{che01}, \citet{lin03}) 
\begin{equation} \label{eqn-cpl}
w(a)=w_0 + w_1(1-a)
\end{equation}
where we also consider the case where $w_1=0$ for values of $w_0$ not
equal to $-1$. 

DNT also consider three more parametrizations labeled Models
1 through 3.  Here we only consider Model 1 where
\begin{equation} \label{eq-mod1}
w(a) = w_f + \frac{w_p-w_f}{1+(a/a_t)^{1/\tau}}
\end{equation}
where $w_f$ is a future value of $w$, $w_p$ is a past value of $w$, $a$ is the scale
parameter, $a_t$ is a transition epoch and $\tau$ characterizes the transition time.
For the purposes of this work Model 1 is considered general enough to represent most
behaviors of $w$.  

\subsection{Constant and Linear Equations of State} \label{ss-cleos}

From equation~\ref{eq-muint} it is obvious that the observed
constraints on $\Delta \mu /\mu$ can be met for any EoS by simply lowering the
value of $\zeta_{\mu}$.  This frees up cosmological parameter space at the 
expense of further limiting the new physics parameter space. We will use our
benchmark $\zeta_{\mu}$ to evaluate the different EOS parameters with the full
knowledge that it is based on the reported value of $\Delta \alpha/ \alpha \approx 10^{-5}$

\subsubsection{Constant EoS models} \label{sss-con}
\begin{figure}
  \vspace{70pt}
\resizebox{\textwidth}{!}{\includegraphics[0in,0in][14.in,3.in]{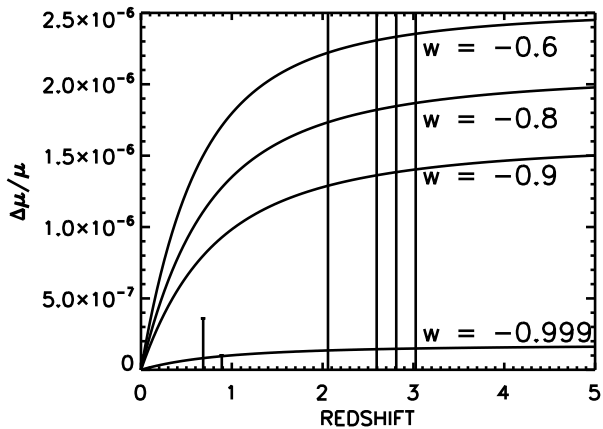}}
  \caption{Values of $\Delta \mu / \mu$ versus redshift for constant value EoS
models with the values labled in the plot. In this and subsequent plots we leave
in the higher redshift constraints even though they are off scale just to show that
they are met.} \label{fig-con}
\end{figure}

As noted at the end of Section~\ref{s-wmu} any deviation of $w$ from $-1$
even if its value is constant will produce a time variation of $\mu$. The magnitude of
the variation depends on the range of the scale factors $a$ where $w$ deviates from $-1$
and the magnitude of the deviation.  Constant EoS models are CPL models with $w_1=0$.
We investigate four different constant EoS models, $w=-0.6,-0.8,-0.9,-0.999$.
Figure~\ref{fig-con} shows the evolution of $\Delta \mu /\mu$ for the constant
EoS cases. Since these models have values of $w$ different from $-1$ for all
scale factors they show significant evolution of $\mu$. Only the $w=-0.999$ 
case matches the low redshift $\Delta \mu /\mu$ constraints which indicates
that most constant EoS models are highly disfavored.
This is consistent with the previous work of \citet{thm13b} that concluded 
that $w$ must be within $0.001$ of $-1$ between a redshift of 0.88582 and
the present day, a time on the order of half of the age of the universe.
Since we denote a change in $\mu$ as a difference between the value of $\mu$
at a given redshift and its value today a $\Delta \mu /\mu$ observation
constrains the evolution of $w$ between the scale factor of the observation
and the scale factor now, taken to be $1$. This is an example of
how a non-geometrical test of the EoS can be very sensitive to models that
would be difficult to discriminate against with a geometric test.

As an example \citet{suz12} used a combination of Hubble Space Telescope and ground based data
to constrain the value of $w$ based on type 1a Supernova, Baryonic Acoustic Oscillations,
Cosmic Microwave Background, and Hubble Constant analysis.  For the case of constant $w$
they found $w=-1.006^{+0.110}_{-0.113}$ for $z<0.5$ and $w=-0.69^{+0.80}_{-0.98}$ for
$0.5<z<1.0$ combining both statistical and systematic errors.  Strictly speaking these
bounds are on the order of 100 less restrictive than the restriction of $w=-1^{+0.001}
_{-0.001}$ found here, however, this is based on a hard limit on $\zeta_{\mu} > 4.0 
\times 10^{-6}$.  At this point there is no appropriate statistical boundary on that limit.

\subsubsection{Linear EoS Models} \label{sss-lin}
\begin{figure}
  \vspace{70pt}
\resizebox{\textwidth}{!}{\includegraphics[0in,0in][14.in,3.in]{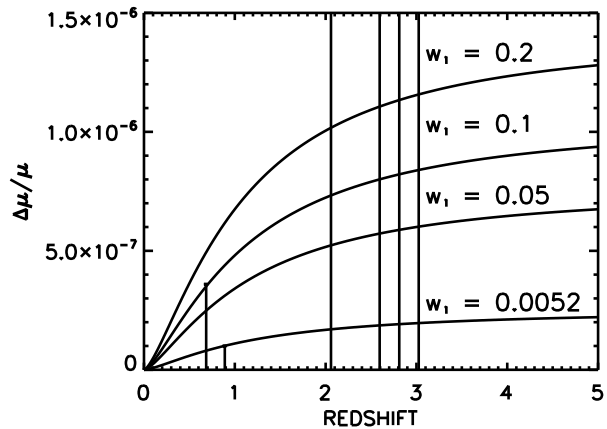}}
  \caption{Values of $\Delta \mu / \mu$ versus redshift for $w_0 = -1$ and varying
$w_1$. The values of $w1$ are labeled in the plot.} \label{fig-lin}
\end{figure}

For linear equations of state we use the simple equation~\ref{eqn-cpl} CPL formulation
with just the two parameters, the current value of $w$, $w_0$ and the slope $w_1$.
In Section~\ref{sss-con} we saw that only models with $w_0$ very close to $-1$
satisfied the constraints so in this section we set $w_0 = -1$ and vary the value 
of $w_1$ to test what range of linear models satisfy the $\Delta \mu /\mu$ constraints.
We chose $w_1$ values of $0.2,0.1,0.05$ and $0.0052$ with the last value the only one that 
satisfies all of the constraints.  Figure~\ref{fig-lin} shows the evolution of $\Delta \mu /\mu$.
 The net result
is that only models with very shallow slopes $(w_1<0.0052)$ and current EOS values very near $-1$ 
satisfy the observational constraints on $\Delta \mu / \mu$. We should note that all
of the linear EOS models that have present day values of $-1$ and negative slopes in
$a$ are crossing the phantom divide, $w < -1$, at the present time. Those with positive
slope have been in the phantom space at earlier times. It would
appear that the parameter space for CPL equations of state is extremely limited other
than the trivial case of $w=-1$.

\citet{suz12} also consider the case of the CPL linear model and find that 
$w_0=-1.046^{+0.179}_{-0.170}$ and $w_1=0.14^{+0.60}_{-0.76}$. Again our formal results
suggest a much tighter bound on both $w_0$ and $w_1$ modulo the validity of the
lower bound on the coupling constant $\zeta_{\mu}$.

\subsection{Freezing and Thawing Equations of State} \label{ss-fteos}
\begin{figure}
  \vspace{70pt}
\resizebox{\textwidth}{!}{\includegraphics[0in,0in][14.in,3.in]{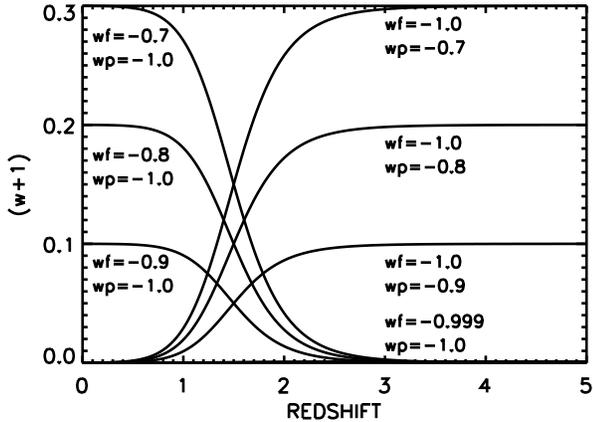}}
  \caption{Values of $(w+1)$ versus redshift for Model 1 with the future, $wf$,
and past, $wp$, values labeled on the plot. Note that the thawing model with 
$w=0.999$ has a future $(w+1)$ value of $0.001$ which is not visible on the scale
of this plot. All of the cases have $\tau=0.1$ and $a_t=0.4$.} \label{fig-m1wpo}
\end{figure}

Freezing equations of state start with an EOS different from $-1$ and approach $-1$
at the present time.  Thawing equations of state start with and EOS very close to
$-1$ and diverge from $-1$ at the present time. To investigate the differences between
the two types of EOS forms we utilize Model 1 of DNT given by equation~\ref{eq-mod1}.
The model has four parameters, $w_p$, the value of $w$ in the past, $w_f$, the value
of $w$ in the future, $\tau$, the transition time, and $a_t$ the transition epoch. 
We chose a value of $\tau = 0.1$ to 
make the transition fast enough to define to regions with well defined asymptotic 
behavior.  The transition epoch is chosen as $a_t = 0.4$ which corresponds to a redshift
of $z=1.5$.  This produces a value of $w$ very near $-1$ in the present epoch for 
freezing equations of state.  Figure~\ref{fig-m1wpo} shows the values of $(w+1)$
versus redshift for a range of values of $w_p$ and $w_f$. 
In all of the cases we have set $\tau=0.1$ and $a_t=0.4$.

It is fairly easy to see how the plots in Figure~\ref{fig-m1wpo} change as the parameters
are varied. Varying the transition epoch $a_t$ moves the transition region along the redshift
axis and varying the transition time $\tau$ changes the steepness of the transition.
\begin{figure}
  \vspace{70pt}
\resizebox{\textwidth}{!}{\includegraphics[0in,0in][14.in,3.in]{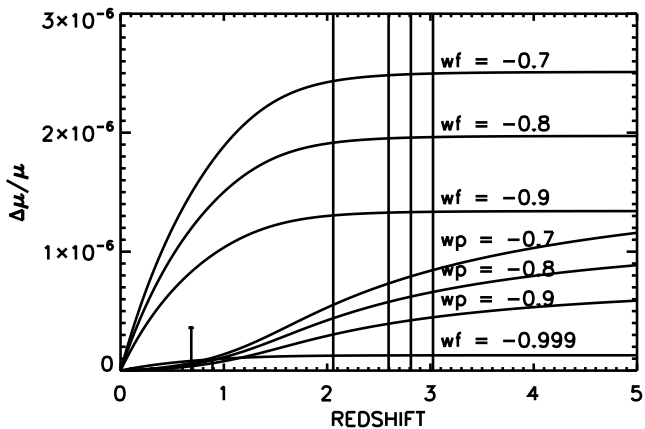}}
  \caption{Values of $\Delta \mu/\mu$ versus redshift for Model 1 with the future, $wf$,
and past, $wp$, values labeled on the plot. The plots labeled with $wf$ values are the
thawing models shown in figure~\ref{fig-m1wpo} and those labeled with $wp$ values are
the freezing models.} \label{fig-m1mu}
\end{figure}

\begin{figure}
  \vspace{70pt}
\resizebox{\textwidth}{!}{\includegraphics[0in,0in][14.in,3.in]{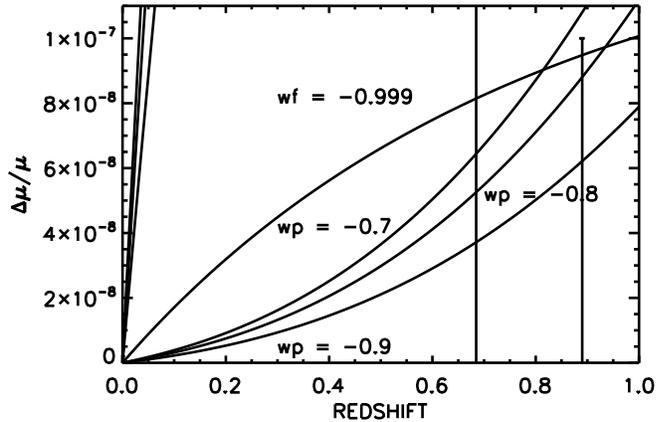}}
  \caption{This is an expanded view of the late time $\mu$ evolution shown in 
Figure~\ref{fig-m1mu} to show the detail of the $\Delta \mu/\mu$ near the most
restrictive constraint at $z=0.88582$.} \label{fig-m1min}
\end{figure}

We again use equation~\ref{eq-muint} and~\ref{eq-omgc} to calculate the evolution of $\mu$
with redshift or scale factor.  Figure~\ref{fig-m1mu} shows the evolution of $\mu$ for
the parameters used in the EOS models shown in Fig.~\ref{fig-m1wpo}.  The evolution of 
$\mu$ shown in Figure~\ref{fig-m1mu} follows the expected trajectories
with thawing EoS models showing significantly more late time evolution due to their
present day deviation from $w=-1$.  Figure~\ref{fig-m1min} shows that all of the freezing
models either meet the constraint at $z=0.88582$ or, in the case of $w_p=-0.7$, miss it
by a very small amount.  These models make it clear that the current much looser constraints
at high redshift from the optical observations allow significant early time evolution of
$w$ as long as the value of $w$ approaches $-1$ by a redshift of $1$ or higher. We can say
that thawing models with significant deviations from $-1$ are strongly disfavored.  The
looser constraints at high redshift can not put the same condition on freezing models. 
Consistent with the results from the linear EoS models the thawing model with a future
value of $w = -0.999$ does meet the low redshift constraint.

\section{Conclusions} \label{s-con}

For a given value of the coupling the deviation of $\mu$ or $\alpha$ depends on the amount of 
time, redshift or scale factor that $w$ deviates from $-1$. This feature strongly disfavors
an EoS that has a constant value not equal to $-1$ as this maximizes the amount of time at a
value of $w$ other than $-1$. The current limit on a constant $w$ is $w=-1 \pm 0.001$ at the
$3\sigma$ level.  This is essentially indistinguishable from $-1$ in geometric tests. Similarly
linear or CPL models are also disfavored at a slightly lower level and are limited to slopes less
than $0.005$ for the case of $w_0=-1$. Any value of $w_0$ other than $-1$ is further disfavored
as the constant term also creates evolution of the constants.

The current data set places much more stringent constraints on thawing models than freezing models
since fundamental constant tests constrain the evolution during the time between the observation
and the present day unlike geometric tests that constrain the evolution that occurred before the
observations.  Since the most stringent constraints are for observations at $z<1$ thawing models 
with significant late time evolution are particularly limited by the data.  Freezing models with
most of their evolution before the epoch of the observations are not as tightly constrained but it
would not be correct to declare either model as more favored or disfavored but rather as less or
more constrained.  All of the model 1 freezing
cases either satisfy or come very close to satisfying the constraints even with past EoSs quite 
deviant from $-1$. The only thawing model 1 to pass the test had the its final value of $w$ at
$w=-0.999$ similar to the constant $w$ case.  However, moving the transition epoch to a later
time would disqualify most of the freezing models.

Modulo the uncertainty on the true lower limit on $\mid \zeta_{\mu} \mid$ the range of possible
EoS models is significantly constrained by the limits on the variation of $\mu$.  In particular the
simple CPL linear EoS model has a very limited range of values which only accommodates slight
deviations from $-1$.  At this point, except for the reported change in the value of $\alpha$
by \citet{web11}, the fundamental constant data are consistent with a cosmological constant
and the standard model of physics.  Higher accuracy measurements at high redshift will be particularly
relevant for putting more significant limits on EoS models in general and on particular cosmological
models.

\section*{Acknowledgments}
The author would like to acknowledge very useful discussions with Carlos Martins on many
aspects of the work present here as well as the many fruitful interactions during this 
conference.

\label{lastpage}

\begin{thebibliography}{99}
\bibitem[\protect\citeauthoryear{Avelino et al.}{2006}]{ave06} Avelino, P.P, Martins, C.J.A.P., 
       Nunes, N.J. \& Olive, K.A. 2006, Phys. Rev. D., 74, 083508
\bibitem[\protect\citeauthoryear{Bagdonaite et al.}{2013}]{bag13} Bagdonaite, J., Jansen, P., 
	Henkel, C., Bethlem, H.L., Menten, K.M. \& Ubachs, W. 2013, Science, 339, 46
\bibitem[\protect\citeauthoryear{Chevallier and Polarski}{2001}]{che01} Chevallier, M. \&
	Polarski, D. 2001, Int. J. Mod. Phys. D, 10, 213
\bibitem[\protect\citeauthoryear{DeFelice, Nesseris \& Tsujikawa}{2012}]{def12} DeFelice ,A.,
	Nesseris, S., \&  Tsujikawa, S. 2012, JCAP05 (2012) 029, DNT
\bibitem[\protect\citeauthoryear{Dutta \& Scherrer}{2011}]{dut11} Dutta, S. \& Scherrer, R.J. 2011,
	Phys. Lett. B, 704, 265
\bibitem[\protect\citeauthoryear{Kanekar}{2011}]{kan11} Kanekar, N. 2011, Ap.J.L., 728, L12
\bibitem[\protect\citeauthoryear{King et al.}{2009}]{kin09} King, J. A., Webb, J. K., Murphy, M. T. \&
	Carswell, R. F. 2009, PRL, 101, 251304
\bibitem[\protect\citeauthoryear{King et al.}{2011}]{kin11} King, J. A., Webb, J. K., Murphy, M.,
        Ubachs, W, \& Webb, J. 2011, MNRAS, 417, 3010
\bibitem[\protect\citeauthoryear{Linder}{2003}]{lin03} Linder, E.V. 2003, Phys. Rev. Lett., 90, 091301
\bibitem[\protect\citeauthoryear{Malec et al.}{2010}]{mal10} Malec, A.L. et al. 2010, MNRAS, 403, 1541
\bibitem[\protect\citeauthoryear{Nunes \& Lidsey}{2004}]{nun04} Nunes, N.J. \& Lidsey, J.E. 2004, Phys 
      Rev D, 69, 123511
\bibitem[\protect\citeauthoryear{Perlmutter et al.}{1999}]{per99} Perlmutter et al. 1999, Ap.J, 517, 565
\bibitem[\protect\citeauthoryear{Riess et al.}{1998}]{rie98} Riess, A.G. et al. 1998, A.J., 116, 1009
\bibitem[\protect\citeauthoryear{Salzano et al.}{2013}]{sal13} Salzano, V., Wang, Y., Sendra, I. \&
	Lazkov, R. 2013, arXiv:1211.1012V2
\bibitem[\protect\citeauthoryear{Suzuki et al.}{2012}]{suz12} Suzuki, N. et al. 2012, Ap.J., 746, 85
\bibitem[\protect\citeauthoryear{Thompson}{2012}]{thm12} Thompson, R.I., 2012, MNRAS Letters, 422, L67
\bibitem[\protect\citeauthoryear{Thompson}{2013}]{thm13b} Thompson, R.I., 2013 MNRAS, 431, 2576
\bibitem[\protect\citeauthoryear{Webb et al.}{2011}]{web11} Webb, J.K., J.A., Murphy, M.T., 
     Flambaum, V.V., Carswell, R.F., \& Bainbridge, M.B. 2011, PRL, 107, 191101-1-5
\bibitem[\protect\citeauthoryear{Weinberg}{1989}]{wei89} Weinberg, S. 1989, Rev. Mod. Phys. 61, 1
\bibitem[\protect\citeauthoryear{Wendt \& Reimers}{2008}]{wen08} Wendt, M. \& Reimers, D. 2008, 
      Eur. Phys. J. ST, 163, 197
\end{thebibliography}
\end{document}